\documentclass[twocolumn,pre, amsmath, amssymb,superscriptaddress]{revtex4}

\usepackage{graphicx}
\usepackage{dcolumn}
\usepackage{bm}
\usepackage{braket}
\usepackage[table]{xcolor}

\usepackage{amsmath}

\newcommand{\beq}{\begin{equation}}
\newcommand{\eeq}{\end{equation}}

\begin{document}


\title{Molecular mechanisms behind anti SARS-CoV-2 action of lactoferrin}

\author{Mattia Miotto}
\affiliation{\footnotesize{Department of
Physics, University of Rome `La Sapienza', Piazzale Aldo Moro, 5,
I00185, Rome, Italy}}
\affiliation{\footnotesize{Fondazione Istituto Italiano di
Tecnologia (IIT), Center for Life Nano Science, Viale Regina Elena
291, I00161 Roma, Italy}} 

\author{Lorenzo Di Rienzo}
\affiliation{\footnotesize{Fondazione Istituto Italiano di
Tecnologia (IIT), Center for Life Nano Science, Viale Regina Elena
291, I00161 Roma, Italy}}

\author{Leonardo B\`{o}}
\affiliation{\footnotesize{Fondazione Istituto Italiano di
Tecnologia (IIT), Center for Life Nano Science, Viale Regina Elena
291, I00161 Roma, Italy}}

\author{Alberto Boffi}
\affiliation{\footnotesize{Department of Biochemical Sciences `A. Rossi Fanelli' Sapienza University, Piazzale Aldo Moro 5, 00185, Rome, Italy
}}

\affiliation{\footnotesize{Fondazione Istituto Italiano di
Tecnologia (IIT), Center for Life Nano Science, Viale Regina Elena
291, I00161 Roma, Italy}}

\author{Giancarlo Ruocco}
\affiliation{\footnotesize{Fondazione Istituto Italiano di
Tecnologia (IIT), Center for Life Nano Science, Viale Regina Elena
291, I00161 Roma, Italy}} 
\affiliation{\footnotesize{Department of
Physics, University of Rome `La Sapienza', Piazzale Aldo Moro, 5,
I00185, Rome, Italy}}

\author{Edoardo Milanetti}
\affiliation{\footnotesize{Department of
Physics, University of Rome `La Sapienza', Piazzale Aldo Moro, 5,
I00185, Rome, Italy}}
\affiliation{\footnotesize{Fondazione Istituto Italiano di
Tecnologia (IIT), Center for Life Nano Science, Viale Regina Elena
291, I00161 Roma, Italy}} \

\begin{abstract}
Despite the huge effort to contain the infection, the novel SARS-CoV-2 coronavirus has rapidly become pandemics, mainly due to its extremely high human-to-human transmission capability, and a surprisingly high viral charge of symptom-less people. While the seek of a vaccine is still ongoing,  promising results have been obtained with antiviral compounds. In particular, lactoferrin is found to have beneficial effects both in preventing and soothing the infection.
Here, we explore the possible molecular mechanisms with which lactoferrin interferes with SARS-CoV-2 cell invasion,  preventing attachment and/or entry of the virus.  
To this aim, we search for possible interactions lactoferrin may have with virus structural proteins and host receptors.  Representing the molecular iso-electron surface of proteins in terms of 2D-Zernike descriptors, we (i) identified putative regions on the lactoferrin surface able to bind sialic acid receptors on the host cell membrane, sheltering the cell from the virus attachment; (ii) showed that no significant shape complementarity is present between lactoferrin and the ACE2 receptor, while (iii) two high complementarity regions are found on the N- and C-terminal domains of the SARS-CoV-2 spike protein, hinting at a possible competition between lactoferrin and ACE2 for the binding to the spike protein.
\end{abstract}

\maketitle

\section{Introduction}

Lactoferrin (Lf) is a versatile glycoprotein, which plays a key role in many biological functions~\cite{redwan2014potential}. In this work, we focus on the Lf as a crucial player in natural immunity, since it has been proposed to play a strong antiviral activity against a wide range of RNA and DNA viruses \cite{waarts2005antiviral,andersen2001lactoferrin,marchetti2004inhibition,drobni2004lactoferrin, puddu1998antiviral}.
Lf is composed of a single chain of about 700 residues folded into two symmetrical lobes. Each lobe possesses a metal-binding site, able to bind iron but also other ions like $Cu^{2+}$, $Zn^{2+}$ and $Mn^{3+}$~\cite{baker2005lactoferrin, Giansanti2016}. 

This protein is present in saliva, tears, seminal fluid, white blood cells, and milk of mammals~\cite{Niaz2019}. From its discovery in 1939~\cite{sorensen1940,groves1960}  lactoferrin has been identified as the most important iron-binding protein in milk. Besides, in recent years, lactoferrin has been found involved in a multitude of biological processes. In fact, despite the name, the iron cargo capacity of Lf is not the prominent activity exerted by this molecule. Instead, it performs antioxidant, anti-inflammatory, and anticancer activities~\cite{caccavo2002, Giansanti2016}, together with a broad antimicrobial action against bacteria and fungi. The latter activity, in particular, is due to Lf's ability to reversibly bind two atoms of iron with high affinity in the presence of bicarbonate. The iron-free form of Lf, apo-lactoferrin (apoLf), deprives bacteria of iron, thus inhibiting their metabolic activities in vivo.

Besides all the aforementioned activities, Lf has been demonstrated to prevent infection of a wide range of diverse viral species~\cite{vanderStrate2001}.

Many viruses make use of glycans, such as sialic acid (SIA) or glycosaminoglycan, like heparan sulfate (HF)  as attachment factors. See, for example,~\cite{LangfordSmith2015} for details on glycans. 
When the contact between the virus particle and these receptors is established, they roll toward their specific viral receptor and subsequently enter the host cell, for instance by
fusing with the host cell membrane~\cite{Raman2016}. 

While the interaction between Lf and HF has been observed~ \cite{Lang2011}, studies on its interaction with sialic acid derivatives are still missing. On the other hand, Lf has been reported to interact with virus structural proteins, S, M, and E~\cite{pmid9223490}. 

In general, depending on the specifics of the virus,  lactoferrin prevents infection of the target cell by either (i) interfering with the attachment factor or (ii) by binding to host cell molecules that the virus uses as a receptor or co-receptor (competition) or (iii) by direct binding to virus particles, as described for herpesvirus~\cite{Harmsen1995}, polio- and rotavirus~\cite{McCann2003,superti2001involvement}, and possibly human immunodeficiency virus~\cite{Puddu1998}. See, for example,~\cite{Berlutti2011} for a more detailed discussion.


While we are writing this article, a novel virus, first observed in the autumn of 2019, has rapidly become pandemic. 
This virus, called SARS-CoV-2, belongs to the coronavirus family and  causes severe acute respiratory syndrome
\cite{huang2020clinical,zhu2020novel},
somewhat similar to those caused by two other coronaviruses, SARS-CoV and MERS-CoV, which crossed species in 2002-2004~\cite{drosten2003identification, ksiazek2003novel} and 2012~\cite{zaki2012isolation}.
In fact, SARS-CoV-2, similarly to SARS-CoV and MERS-CoV, attacks the lower respiratory system, thus provoking viral pneumonia. However, this infection can also lead to effects on the gastrointestinal system, heart, kidney, liver, and central nervous system \cite{prompetchara2020immune, su2016epidemiology, zhu2020novel}. 

As for SARS-CoV~\cite{li2005structure,li2008structural,li2005receptor}, recent in vivo experiments confirmed that also SARS-CoV-2 cell entry is mediated by high-affinity interactions between the receptor-binding domain (RBD) of the virus S glycoprotein and the human-host Angiotensin-converting enzyme 2 (ACE2) receptor~\cite{zhou2020pneumonia}. The spike protein is located on the virus envelope and promotes the host attachment and fusion between the viral and cellular membrane. \cite{graham2010recombination, kuo2000retargeting}. Structural studies determined the structures of such protein both in free form and bound to ACE2~\cite{yan2020structural}. 
Further studies investigate the possible interaction of SARS-CoV-2 to   sialic acids~\cite{milanetti2020silico,vandelli2020structural,Liu2020,robson2020bioinformatics} or heparan surfate receptors~\cite{Liu2020}, both considered involved in SARS-CoV-2  as well as in other coronavirus infections~\cite{ schwegmann2006sialic, Lang2011, tortorici2019structural, hulswit2019human,LangfordSmith2015}.

\begin{figure*}[t]
\centering
\includegraphics[width = \textwidth]{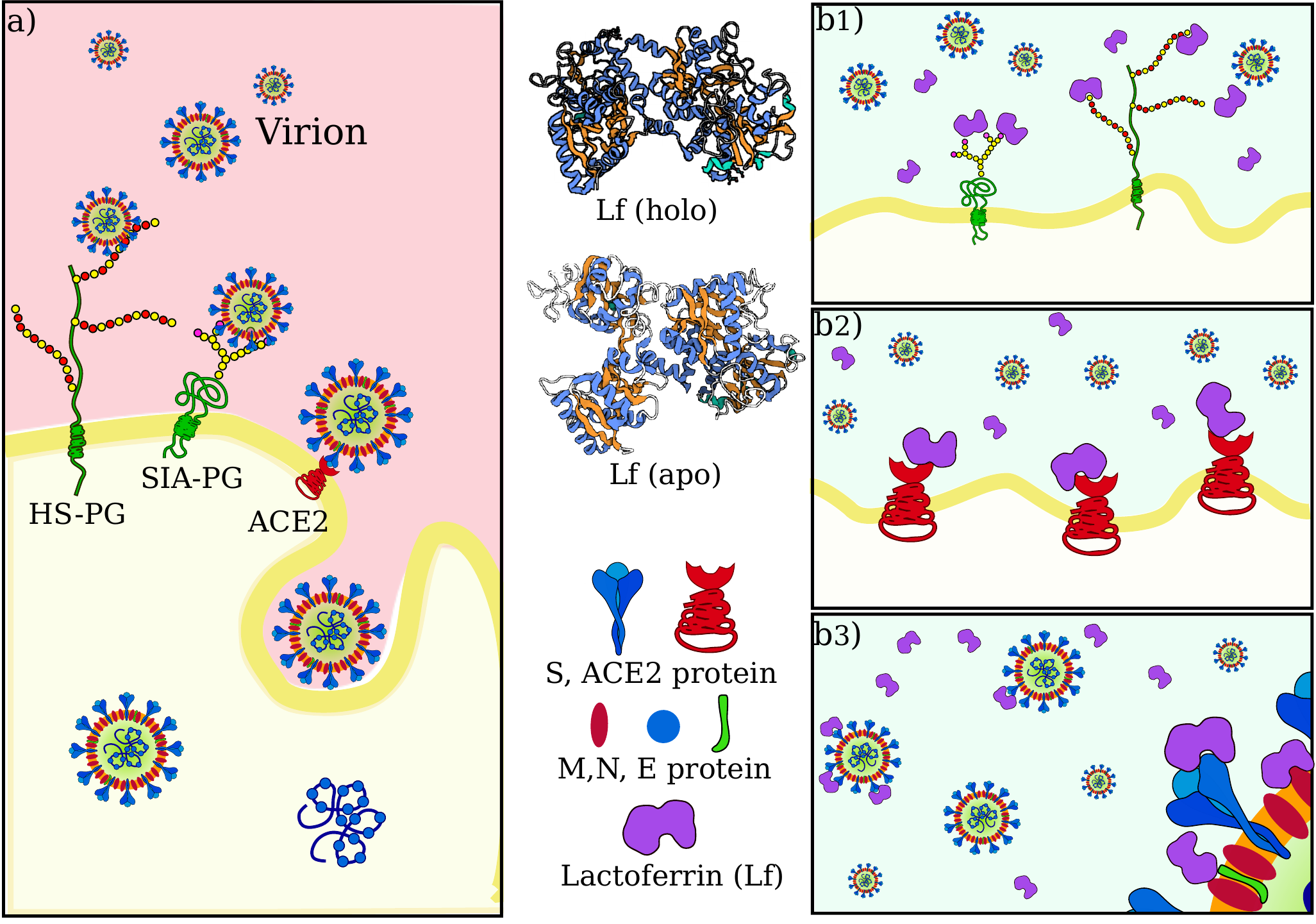}
\caption{\textbf{SARS-CoV-2 attachment and entry to host cell in physiological condition and possible actions of lactoferrin.}
\textbf{a)} Sketch of SARS-CoV-2 initial interactions with the host cell. Sialoside (SIA) and heparan sulfate (HS) glycan chains present on glycoproteins (PG) of the cell membrane are thought to facilitate the attachment of the virion to the cell surface.  This favors the establishment of an interaction between the virus spike protein and the ACE2 receptor, which starts the internalization of the virus in the host cell.
\textbf{b)} Human lactoferrin has been found to play an antiviral action against SARS-CoV-2 infection although it is not clear whether this action consists in (1) competition for the binding with glycan chains, and/or (2) competition for binding ACE2 receptor and/or (3) direct interaction with one of the proteins in the virion envelope, i.e. with  S, M or E proteins.  }
\label{fig:scheme}
\end{figure*}

While the use of Lf as an antiviral against previous coronaviruses infections has been poorly investigated, except for~\cite{Lang2011}, where evidence of an effect in the attachment process is shown, new studies are proving the antiviral effect of Lactoferrin against the novel SARS-CoV-2 infection. 
In particular,  administration  of a liposomal formulation of Lf to a significant sample of  Covid-19 positive patients, has been shown to provide an immediate beneficial effect~\cite{serranoliposomal}.

Here, we computationally investigate the possible molecular mechanisms behind the observed antiviral action of lactoferrin. In particular, we make use of a recently developed computational protocol based on the 2D Zernike Polynomials, able to rapidly characterize the shape conformation of given protein regions~\cite{milanetti2020silico}. In this framework using a simple pairwise distance, it is possible to evaluate the similarity between 2 protein pockets or the shape complementarity between the binding regions of 2 interacting proteins.

In order to assess whether lactoferrin could influence the attachment factors, we investigated the ability of Lt to bind to sialic acid (SIA) or heparan sulfate (HS) receptors~\cite{LangfordSmith2015}, both considered involved in SARS-CoV-2 infection~ \cite{milanetti2020silico,vandelli2020structural, Liu2020,robson2020bioinformatics} as well as to other coronavirus infections~\cite{ schwegmann2006sialic, Lang2011, tortorici2019structural, hulswit2019human}. 

We moreover checked for a possible direct interaction between LF and ACE2 receptor, which could inhibit the interplay between spike and ACE2 binding necessary to the virus infection. 

Finally, we investigate the possible interaction between Lf and the 3  proteins present on the SARS-CoV-2 membrane, i.e. the spike (S), membrane (M), and envelope (E) proteins.

\begin{figure*}[t]
\centering
\includegraphics[width = \textwidth]{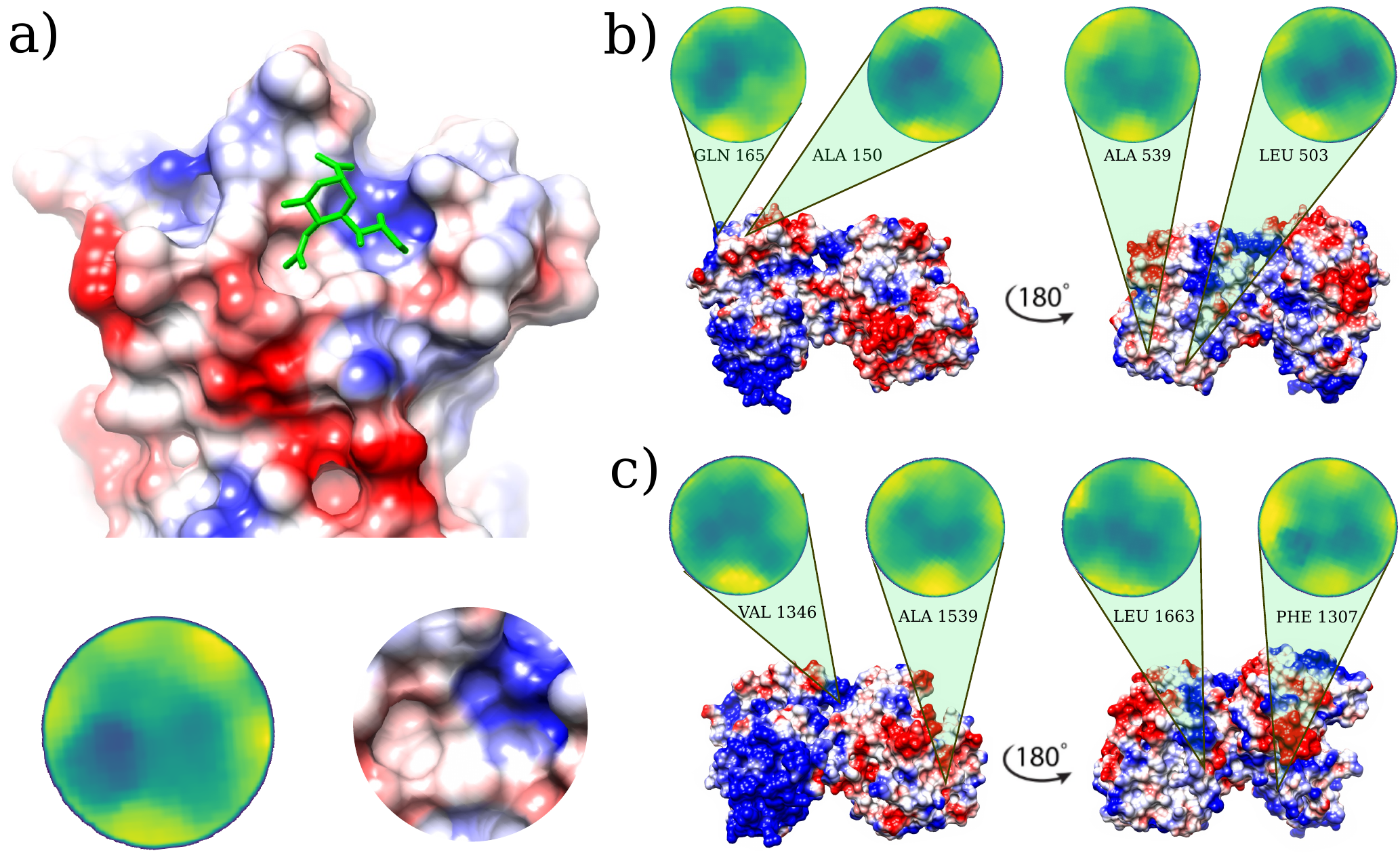}
\caption{\textbf{Putative sialic-binding regions on human lactoferrin.}  \textbf{a)} Cartoon representation of the sialic-acid binding region of MERS-CoV (PDB id: 6Q04) and its representation in the Zernike disk (left) and local Coulombic surface (right). \textbf{b)} Four most complementar regions on holo human lactoferrin (PDB id: 1LFG) obtained comparing the lactoferrin molecular surface with the sialic acid binding region of MERS-CoV. \textbf{c)} Same as in b) but for the apo human lactoferrin (PDB id: 1CB6). Molecular surfaces are colored according to the Coulombic potential.}
\label{fig:2}
\end{figure*}

\section{Results}

The entry of the virus inside the host cells requires the occurrence of a sequence of molecular interactions. 
Sialoside (SIA) and/or heparan sulfate (HF) chains mediate the attachment of the virion to the cell surface. Once in the proximity of the cellular receptor (ACE2), SARS-CoV-2 spike protein binds to the receptor and initiates the internalization process. Figure~\ref{fig:scheme}a shows a sketch of the mechanism.
The observed antiviral action of Lf may consist of interference in one or more of those steps. We thus investigate in the next three sections the possibility of direct binding between lactoferrin and SIA, ACE2, and SARS-CoV-2 Spike protein, respectively (see Figure~\ref{fig:scheme}b).

\subsection{Interaction with SIA}

Possible binding regions for sialic acid, the terminal molecule of the SIA receptors on human lactoferrin are investigated on the basis of  the procedure described in~\cite{milanetti2020silico}, i.e. we select the portion of the molecular surface of the MERS-CoV spike protein in interaction with sialic acid, experimentally solved in~\cite{Park2019} (Figure~\ref{fig:2}a), and we search for similar patches on the Lf molecular surface.

Within the same strategy, we search for similar patches on the surface of human lactoferrin.  

In the 2D Zernike framework, the geometrical shape of a protein surface patch is compactly summarized in a set of ordered numerical descriptors, whose number -  121 in our case - modulate the detail of the description. Dealing with ordered numerical descriptors, the comparison between different protein patch can be performed with a Euclidean distance. 

Figure~\ref{fig:2}b and c show the four most geometrically similar patches identified in the Apo and Holo form of Lf. An \textit{a posteriori} check of the electrostatic potential on the patches,  allows us to select only some of the possible solutions identified based on the shape comparison analysis, i.e. the ones having also a similar electrostatic surface with the  SIA binding site on the MERS-CoV spike.

The region on Lf surface identified as the most similar to the MERS region interacting with sialic acid, both in shape and in electrostatics, is the one centered on VAL 346.

\subsection{Interaction with ACE2}

\begin{figure}[t]
\centering
\includegraphics[width = \columnwidth]{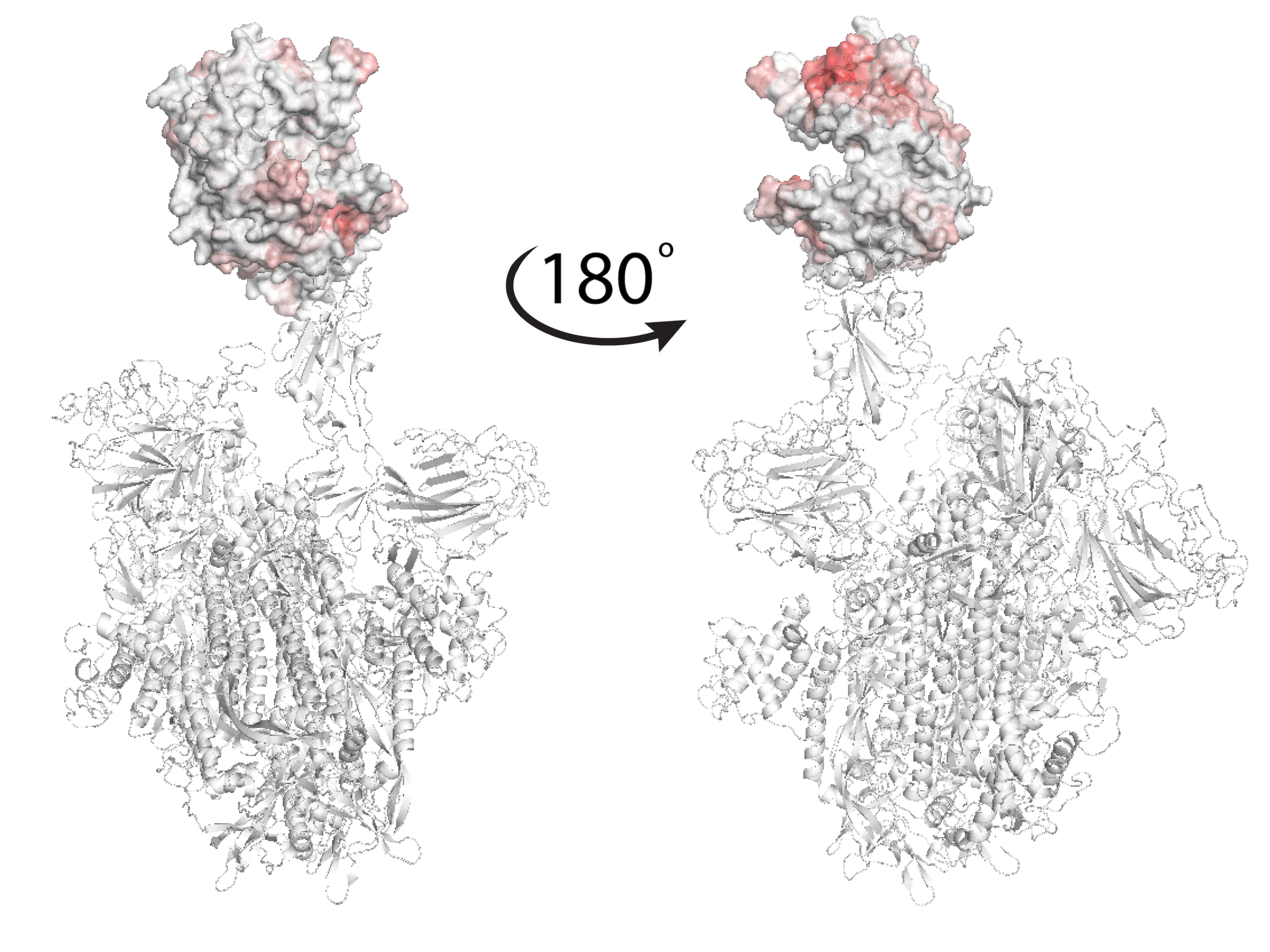}
\caption{\textbf{Analysis of the binding between lactoferrin and the ACE2 receptor.}  
Molecular representation of the experimentally solved complex of the ACE2 receptor with SARS-CoV-2 spike protein. The ACE2 molecular surface is colored according to its binding propensity to bind lactoferrin regions. Dark red indicates high binding propensity while white means no interaction.}
\label{fig:3}
\end{figure}

To check whether the action of lactoferrin can be ascribed to a competition with the virion spike proteins in binding directly the ACE2 receptor (Figure~\ref{fig:scheme}), we performed a blind search of the molecular surfaces of both ACE2 and human Lf to identify possible binding regions having a meaningful shape complementarity. Under this hypothesis, if the interaction between the Lf and the ACE2 receptor occurs, Lf could hinder the molecular binding between the spike protein of SARS-CoV-2 and the corresponding ACE2 receptor. 
Figure~\ref{fig:3} shows the molecular surface of the ACE2 receptor colored according to its propensity to bind regions of the Lf protein. The redder the region, the greater the shape complementarity between that region and another one found on the surface of the putative molecular partner, i.e. holo lactoferrin. 
As one can see from Figure~\ref{fig:3}, a complementary region is indeed present, however, it is located far from the binding site of the spike (grey in the figure) and in a part of ACE2 that looks toward the membrane. To better visualize the result, we have represented two points of view of the binding between spike and ACE2, one rotated 180$^{o}$ with respect to the other. 
On the other hand, we can see that the ACE2 region interacting with the SARS-CoV-2 spike protein has no low shape complementarity with Lf regions.

\subsection{Interaction with virion membrane proteins}

A third possible mechanism at the basis of the observed antiviral activity of Lf could be ascribed to a direct interaction with the membrane proteins present on the virion envelope. In particular, SARS-CoV-2 presents three different kinds of proteins on its membrane, i.e. S, M, and E proteins~\cite {Seah2020, Bianchi2020}. While the 3D structure of the S protein has been determined - even if some loop regions in the S1 sub-unit are not solved - unfortunately, no structures are available for the E and M proteins.
Thus, the molecular surfaces of Lf were compared with those of the
three proteins in order to check whether an interaction with lactoferrin is possible.
For this analysis, we adopted the same computational procedure used in the previous paragraph. 

For both S, M, and E proteins, we sample their whole molecular surface and compared all the possible patches with those of Lf. In this way, all molecular surfaces, both membrane proteins, and Lf ones are colored according to the corresponding binding propensity.

\begin{figure*}[t]
\centering
\includegraphics[width = \textwidth]{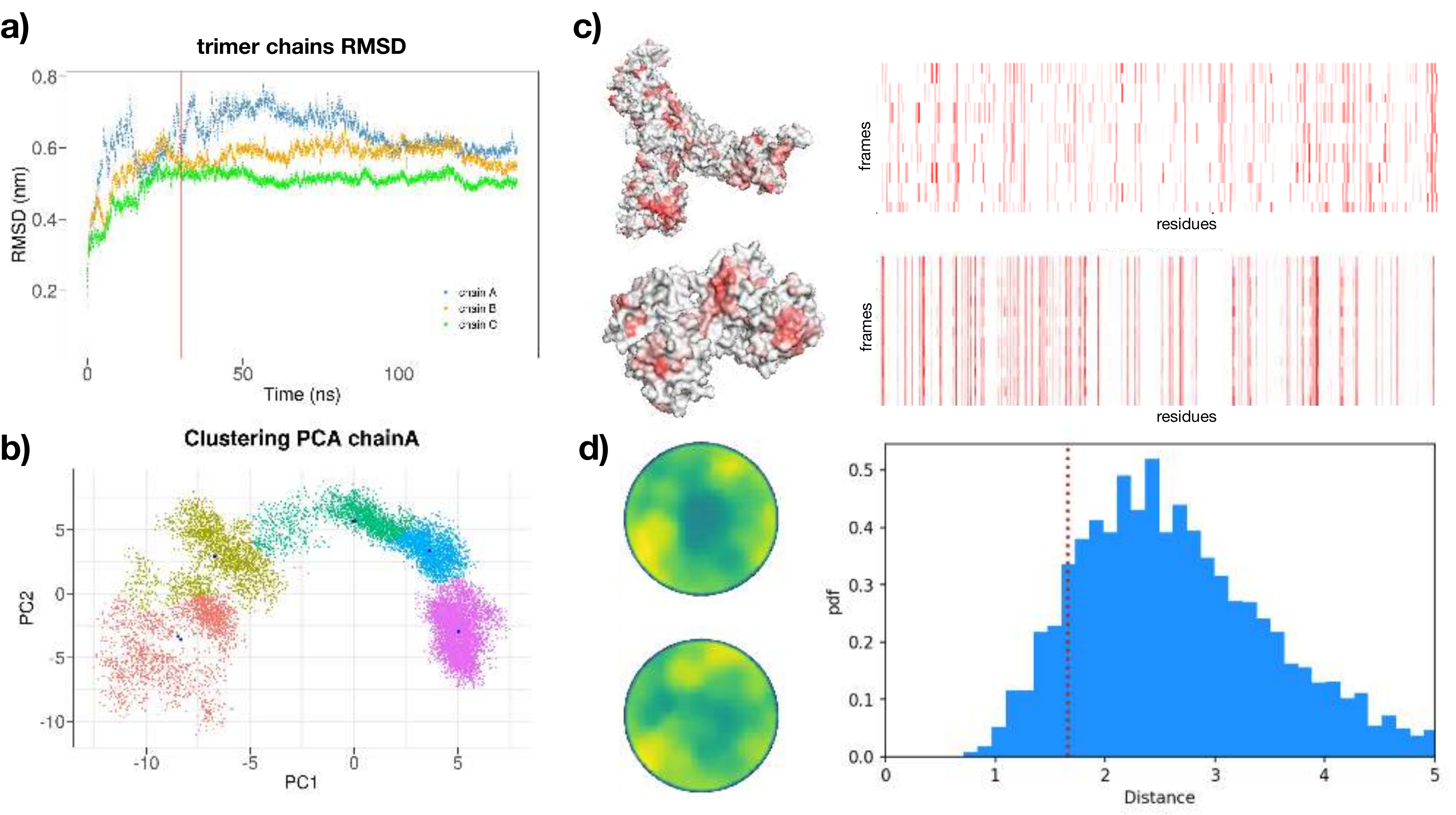}
\caption{\textbf{Possible interaction between SARS-CoV-2 spike protein and human lactoferrin.}  \textbf{a)} Root mean square displacement as a function of time of the SARS-CoV-2 spike trimer as provided by the  molecular dynamics simulation.\textbf{b)} Clustering analysis of the trimer's A chain in the plane of the two principal components of a PCA analysis over the MD configurations.  Five regions of major variability of the chain are identified. \textbf{c)} Binding propensity computed from the Zernike descriptors between the 15 most variable conformations of the chains of the SARS-CoV-2 Spike protein and human lactoferrin. Each residue of the proteins is colored from white to red according to its increasing shape complementarity with the partner.
\textbf{d)} Comparison between the complementarity score (red dashed line) of the best binding site (Zernike disks) and the distribution of complementarity scores belonging to 4600 binding regions of experimental complexes.}
\label{fig:4}
\end{figure*}

E and M presented a possible region of interactions located in the intra-membrane region (data not shown). The most robust and relevant result of this analysis regards the compatibility between the spike and lactoferrin. According to our findings, Lf presents two regions of high complementarity with one portion of the C-terminal domain of the spike S1 subunit and another located in the N-terminal part.

To test the reliability of the found signals, we performed a molecular dynamics simulation of the spike trimer (see Methods for details) and sampled 5 configurations for each of the three chains at equilibrium (see Figure~\ref{fig:4}a).

In particular, for each chain, we performed a Principal Component Analysis (PCA) on the frames of the dynamics and thus projected them on the plane identified by the two principal components. Upon clustering these points we obtain 5 subgroups. For each subgroup, we extract the centroidal configuration. Since distant points in the PCA plane correspond to the different 3D structure, picking one point from each identified  cluster assured that the selected configurations have high structural differences between them in the explored configuration space  (see Figure~\ref{fig:4}b).

Remarkably, repeating the blind search for complementarity regions on the 15 surfaces of the extracted spike monomers, we found conservation of the signal over the conformational noise. Figure~\ref{fig:4}c) shows the identified regions and their conservation in the different sampled frames.

Among the found regions, the one involving the spike C-terminal domain is the one with higher shape complementarity. 
In particular, according to our method, human lactoferrin could interact with the spike protein with the surface region centered in residue~ALA 539. Alternatively, the spike protein may interact with the Lf protein using a molecular patch centered in the residue~PHE 490. 
It must be pointed out that the complementarity achieved by these patches is comparable with those of experimentally solved complexes. Indeed, analyzing the shape complementarity of over 4600 x-ray protein-protein complexes (see Methods for details), we have the distribution shown in Figure \ref{fig:4}d, where the lower the distance the higher the complementarity of the binding region. Red dotted line shows the complementarity found between the spike and Lt. As it is evident the proposed patch is characterized by values of shape complementarity typical of experimental complexes.

Finally, to further support this result with an independent and external methodology to our approach, we performed a completely blind molecular docking analysis between the spike protein and the Lf protein. To this end, Zdock server was used as a state of the art of molecular docking software~\cite{Chen2003}. As per default settings, only the first 10 docking poses have been selected and the predicted contacts analyzed.
In particular, five out of ten poses show bindings involving the spike region involve the region we found with our protocol when the residues are defined in contact if their C-alpha atoms have a distance less than 8~$\AA$.

\section{Discussion}

In the last decade, the Zernike formalism has been widely applied for the characterization of molecular surfaces~\cite{di2017superposition, daberdaku2019antibody, kihara2011molecular,  di2020quantitative, venkatraman2009protein}.

Very recently,  we developed a new representation, based on the 2D Zernike polynomials, which allows an extremely efficient, fast, and completely unsupervised description of the local geometrical shape, allowing for easy comparison between different regions of molecules.
Through this compact description, it is possible both to analyze the similarity between 2 different regions - suggesting, for example, a similar ligand for binding regions - and to study the complementarity between interacting surfaces~\cite{milanetti2020silico}.

Here, we used our novel method to shed some light on the molecular mechanisms ruling the observed antiviral action human lactoferrin exerts against SARS-CoV-2 infection~\cite{serranoliposomal}.

In particular, we focused on the early stages of the infection, i.e. the attachment and entry of the virus to the host cell, when lactoferrin can interfere with the virus-host interaction without the need to be internalized in the cell.
We thus tried to establish whether lactoferrin could compete with the virus in binding to sialic acid, the sticky end of sialoside chains, which has been suggested to mediate the attachment of SARS-CoV-2 to the host cell. Interestingly, comparing the binding region of sialic acid in the MERS-CoV coronavirus with patches on the lactoferrin surface, we found possible spots on both apo and holo forms of Lf, which could compete in forming low affinity but high avidity interactions.

We then proceeded to test the hypothesis of an interaction between lactoferrin and the primary SARS-CoV-2 protein receptor, ACE2. A blind search for complementarity regions highlighted a hot-spot in a region that in physiological conditions is oriented toward the membrane, while no significative complementarity is present in the ACE2 region involved in the interaction with the virus spike protein.
At last, we analyzed the three membrane proteins on the virus envelope, i.e. the E, M, and S ones. Similarly to ACE-2, both E and M presented possible interacting regions in portions of the surface, that are buried in the virion membrane under normal conditions (data not shown). On the other hand, the spike protein showed two main hot spots, one in the N-terminal domain of the S1 subunit and another in the C-terminal one. 
Those two regions are robust to molecular noise, as the signal endures using different configurations sampled from a molecular dynamics simulation and each of the three chains of the trimer. 
Notably, the most complementary region is the one in the C-terminal region,  the one involved in the spike-ACE2 interaction. Thus our finding suggests a possible competition between ACE2 and lactoferrin for the binding of the SARS-CoV-2 spike, which may explain the observed antiviral action.

\section{Materials and Methods}

\subsection{Datasets}
The protein, whose structures are analyzed in this paper are:
\begin{itemize}
\item Human lactoferrin, in the apo (PDB id: 1CB6) and holo (PDB id: 1LFG) forms.
\item ACE2, in its apo state (PDB id: 1R42).
\item SARS-CoV-2 S protein, modeled using I-Tasser~\cite{Roy2010}.
\item SARS-CoV-2 M protein, modeled using I-Tasser~\cite{Roy2010}.
\item SARS-CoV-2 E protein, modeled using I-Tasser~\cite{Roy2010}.
\end{itemize} 

To set a reference for the measured complementarities, a dataset of protein-protein complexes experimentally solved in x-ray crystallography is taken from~\cite{gainza2020deciphering}.
We only selected pair interactions regarding chains with more than 50 residues. The Protein-Protein dataset is therefore composed of 4605 complexes. For each complex, the binding region is identified as the portions of the two protein molecular surfaces distant less than 3 $\AA$.

\subsection{Computation of molecular surfaces}   
For each protein of the dataset (x-ray structure in PDB format~\cite{berman2003protein}),  we use DMS~\cite{richards1977areas} to compute the solvent-accessible surface, using a density of 5 points per $\AA^2$ and a water probe radius of 1.4 $\AA$. The unit normals vector, for each point of the surface, was calculated using the flag $-n$.

\subsection{Patch definition and complementarity evaluation}

A molecular surface is represented by a set of points in the three-dimensional space. 
We define a surface patch,  as the group of points that fall within a  sphere of radius $R_s = 6 \AA$, centered on one point of the surface.
Once the patch is selected, 
\begin{itemize}
\item we fit a plane that passes through the points and reorient the patch in such a way to have the z-axis perpendicular to the plane and going through the center of the plane.
\item we define the angle $\theta$ as the largest angle between the perpendicular axis and a secant connecting a given point $C$ on the z-axis to any point of the patch. $C$ is then set in order that $\theta=45^\circ$. $r$ is the distance between $C$ and a surface point. \item build a square grid and associate each pixel with the mean $r$ of the points inside it. This 2D function can be expanded on the basis of the Zernike polynomials (see next section).
\end{itemize}

Once a patch is represented in term of its Zernike descriptors, the similarity between that patch and another one can be simply measured as
the Euclidean distance between the invariant vectors.
The relative orientation of the patches before the projection in the unitary circle must be considered. In fact, if we search for similar regions we must compare patches that have the same orientation once projected in the 2D plane, i.e. the solvent-exposed part of the surface must be oriented in the same direction for both patches, for example as the positive z-axis. If instead, we want to assess the complementarity between two patches, we must orient the patches contrariwise, i.e. one patch with the solvent-exposed part toward the positive z-axis (`up') and the other toward the negative z-axis (`down').

\subsection{2D Zernike polynomials and invariants}

Each function of two variables, $f(r,\phi)$ (polar coordinates) defined inside the region $r < 1$ (unitary circle), can be decomposed in the Zernike basis as 
\beq
f(r,\phi) = \sum_{n=0}^{\infty} \sum_{m=0}^{m=n} c_{nm} Z_{nm}
\eeq

with 

\begin{multline}
c_{nm} =  \frac{(n+1)}{\pi} \braket{Z_{nm}|f} =\\= \frac{(n+1)}{\pi} \int_0^1 dr r\int_0^{2\pi} d\phi Z_{nm}^*(r,\phi) f(r,\phi).
\end{multline}

being the expansion coefficients, while the complex functions, $Z_{nm}(r, \phi)$ are the    Zernike polynomials. 
Each polynomial is composed by a radial and an angular part,

\beq
Z_{nm} = R_{nm}(r) e^{im\phi}.
\eeq

where the radial part for any  $n$ and $m$, is given by

\beq
R_{nm}(r) = 
\sum_{k= 0}^{\frac{n-m}{2}} \frac{(-1)^k (n-k)!}{k!\left(\frac{n+k}{2} - k\right)!\left(\frac{n-k}{2}-k\right)!} r^{n-2k}
\eeq

Since  for each couple of polynomials, the following relation holds

\beq
\braket{Z_{nm}|Z_{n'm'}} = \frac{\pi }{(n+1)}\delta_{nn'}\delta_{mm'}
\eeq

the complete set of  polynomials forms a basis and 
knowing the set of complex coefficients, $\{ c_{nm} \}$ allows for a univocal reconstruction of the original image (with a resolution that depends on the order of expansion, $N = max(n)$). 
Since the modulus of each coefficient ($z_{nm} = |c_{nm}|$) does not depend on the phase, i.e. it is invariant for rotations around the origin of the unitary circle, the shape similarity between two patches can be assessed by comparing the Zernike invariants of their associated 2D projections.
In particular, we  measured the similarity between patch $i$ and $j$ as
the Euclidean distance between the invariant vectors, i.e.
\beq 
d_{ij} = \sqrt{\sum_{k=1}^{M=121} (z_i^k - z_j^k)^2}
\eeq


\subsection{Molecular dynamics simulations}

The starting structure of the SARS-CoV-2 spike trimeric complex was taken from the model structure proposed by the I-Tasser server~\cite{Roy2010}.  
All steps of the simulation were performed using Gromacs 2019.3~\cite{gromacs}.
Topologies of the system were built using the CHARMM-27 force field~\cite{charmm}.
The protein was placed in a dodecahedric simulative box, with periodic boundary conditions, filled with 131793 TIP3P water molecules~\cite{Jorgensen1983}. 
We checked that each atom of the trimer was at least at a distance of 1.1 nm from the box borders.
The addition of 3 sodium counterions rendered the systems electroneutral.
The final system,  consisting of 448572 atoms, was first minimized with 2064 steps of steepest descent. 
Relaxation of water molecules and thermalization of the system in NVT and NPT environments were run each for 0.1 ns at 2 fs time-step. 
The temperature was kept constant at 300 K with v-rescale algorithm\cite{vrescale}; the final pressure was fixed at 1 bar with the Parrinello-Rahman algorithm\cite{parrinello} which guarantees a water density of 1004 $\mathrm{kg / m^3}$, close to the experimental value. 
LINCS algorithm\cite{lincs} was used to constraint h-bonds.

Finally, the systems were simulated with a 2 fs time-step for 140 ns in periodic boundary conditions, using a cut-off of 12 $\AA$ for the evaluation of short-range non-bonded interactions and the Particle Mesh Ewald method~\cite{Cheatham1995} for the long-range electrostatic interactions.



\end{document}